\documentclass[10pt,conference,compsocconf,letterpaper]{IEEEtran}
\pdfoutput=1
\usepackage{amsmath}
\usepackage{graphicx}
\usepackage[noend]{distribalgo}
\usepackage{algorithm}
\usepackage{amssymb}
\usepackage{color}
\usepackage[draft]{fixme}
\usepackage{multirow}
\usepackage{afterpage}
\usepackage{soul}
\newcommand{\sizefactor}{0.9}

\begin{document}

\title{\mbox{Rethinking State-Machine Replication for Parallelism}}

\author{
    \IEEEauthorblockN{Parisa Jalili Marandi, Carlos Eduardo Bezerra, Fernando Pedone}
    \IEEEauthorblockA{University of Lugano, Switzerland}}

\maketitle

\begin{abstract}
State-machine replication, a fundamental approach to designing fault-tolerant services, requires commands to be executed in the same order by all replicas. 
Moreover, command execution must be deterministic: each replica must produce the same output upon executing the same sequence of commands. 
These requirements usually result in single-threaded replicas, which hinders service performance. 
This paper introduces Parallel State-Machine Replication (P-SMR), a new approach to parallelism in state-machine replication.
P-SMR scales better than previous proposals since no component plays a centralizing role in the execution of independent commands---those that can be executed concurrently, as defined by the service.
The paper introduces P-SMR, describes a ``commodified architecture" to implement it, and compares its performance to other proposals using a key-value store and a networked file system.
\end{abstract}

\section{Introduction}
\label{sec:intro}

State-machine replication (SMR) is a fundamental approach to designing fault-tolerant services~\cite{Lam78, Sch90}.
By replicating the servers, the service remains available for clients even if some replicas fail.
Once servers are replicated, consistency among the replicas must be ensured.
State-machine replication achieves strong consistency by regulating how client commands are propagated to and executed by the replicas: 
(i)~every nonfaulty replica must receive every command; 
(ii)~replicas must agree on the order of received and executed commands; and
(iii)~the execution of commands must be deterministic (i.e., a command's changes to the state and results depend only on the replica's state and on the command itself).

State-machine replication is a technique to improve the availability of a service, not its performance.
In fact, a single-server implementation of a service will likely outperform its replicated counterpart since the single server can benefit from concurrency while the replicated servers will be typically sequential.
Executing commands sequentially is a serious performance limitation in modern processors, which are essentially parallel (i.e., equipped with multiple processing units, interconnected through multiple network interfaces).
Nevertheless, rendering state-machine replication parallel is challenging.

In state-machine replication, upon executing the same sequence of commands replicas evolve through the same sequence of states and produces the same results.
It has been early observed, however, that replicas do not need to pass through the same state sequence to produce the same responses~\cite{Sch90}.
This is the case of commands that access disjoint variables: if commands do not contend for shared variables, replicas can execute them in parallel (i.e., concurrently).
Some previous works have build on this observation to introduce parallelism in state-machine replication~\cite{KWQCAD2012, KD2004}.
As we present in more detail in the next sections, the idea is to deliver commands across replicas in total order but allow them to execute concurrently when possible, as observed above.

In this paper, we seek solutions that not only allow parallelism in state-machine replication but also scale with the hardware resources available at processors (e.g., processing units, network interfaces).
The last requirement has two important consequences.
First, we should avoid solutions that rely on a single component or on a fixed set of components, an obvious potential performance bottleneck.
Second, we must accommodate parallelism in the execution of commands and in the ordering of these commands.
Failing to address these issues will result in limited performance improvements, as more services run at ``main-memory speed" and are deployed on processors equipped with an ever-increasing number of processing units.

Parallel State-Machine Replication (P-SMR), the approach we present in this paper, fulfills the requirements above: it introduces parallelism in the execution of commands and in the protocol used to order these commands; 
additionally, in the most common cases performance scales with replicas' hardware resources.
Similarly to previous proposals, P-SMR exploits service semantics to determine when commands can execute concurrently and when serial execution is needed.
This is captured by the notion of dependency between commands: two commands are deemed dependent if they cannot execute concurrently.
P-SMR is optimized for independent commands, when concurrency is possible.
Services whose state is mostly read (e.g., name services) or can be partitioned so that most commands fall in one partition or another but rarely in both (e.g., file systems) can benefit from P-SMR.

We have implemented P-SMR and compared its performance to other state-machine replication approaches using two services, a key-value store and a networked file system.
In summary, we have found that for independent commands, P-SMR outperforms classical state-machine replication by a factor of more than 3 and other approaches by a factor of more than 2.
Moreover, P-SMR scales performance with the number of cores in a processor.

This paper makes the following contributions:
\begin{itemize}
\item It introduces a novel approach to parallel state-machine replication (P-SMR) that scales performance with the number of processing units in a replica when commands are independent.
\item It describes a ``commodified architecture" for state-machine replication and shows how replicated services built based on this architecture can seamlessly use P-SMR.
\item It shows how P-SMR can be used to boost the performance of two highly available services, a key-value store and a networked file system.
\item It details a prototype of P-SMR, assesses its performance, and compares it to other state-machine replication approaches.
\end{itemize}

The remainder of the paper is structured as follows.
Section~\ref{sec:model} describes our system model and assumptions.
In Section~\ref{sec:background}, we present an architecture for state-machine replication and motivate the need for P-SMR.
In Section~\ref{sec:psmr}, we introduce Parallel State-Machine Replication.
Section~\ref{sec:applic} illustrates how P-SMR can be used with two services.
Section~\ref{sec:imple} discusses the implementation of our highly available services using different strategies.
Section~\ref{sec:evaluation} contains a performance evaluation of these systems.
Section~\ref{sec:rwork} surveys related work and Section~\ref{sec:final} concludes the paper.

\section{System model and assumptions}
\label{sec:model}

We assume a distributed system composed of interconnected processes. 
There is an unbounded set $C = \{ c_1, c_2, ... \}$ of \emph{client} processes and a bounded set $S = \{ s_1, s_2, ..., s_n \}$ of \emph{server} processes. 
The system is asynchronous: there is no bound on messages delays and on relative process speeds.
We assume the crash failure model and exclude malicious and arbitrary process behavior (e.g., no Byzantine failures). 
Processes are either \emph{correct}, if they never fail, or \emph{faulty}, otherwise. 
We assume $f$ faulty servers, out of $n = f+1$ servers.

Processes communicate by message passing, using either one-to-one or one-to-many communication. 
One-to-one communication is through primitives send$(m)$ and receive$(m)$, where $m$ is a message. 
If sender and receiver are correct, then every message sent is eventually received. 
One-to-many communication is based on atomic multicast.
%
Atomic multicast is defined by the primitives \emph{multicast}$(\gamma, m)$ and \emph{deliver}$(m)$, where $\gamma \subseteq S$ is a group of destinations.
Let relation $<$ be defined such that $m < m'$ iff there is a process that delivers $m$ before $m'$.
Atomic multicast ensures that 
(i)~if a server delivers $m$, then all correct servers in $\gamma$ deliver $m$ \emph{(agreement)}; and
(ii)~relation $<$ is acyclic \emph{(order)}.
The order property implies that if $s$ and $r$ deliver messages $m$ and $m'$, then they deliver them in the same order.

Atomic multicast is typically available to applications as a library (Figure~\ref{fig:architecture}) and implemented using one-to-one communication and additional system assumptions~\cite{CT96,Lam98} (see Section~\ref{sec:imple}).
The replication protocols we consider in the paper use atomic multicast as a ``black box" and do not explicitly use these assumptions.

\section{Background and motivation}
\label{sec:background}

State-machine replication renders a service fault-tolerant by replicating the server and coordinating the execution of client commands among the replicas~\cite{Lam78, Sch90}.
The service is defined by a state machine and consists of \emph{state variables} that encode the state machine's state and a set of \emph{commands} that change the state (i.e., the input).
The execution of a command may (i)~read state variables, (ii)~modify state variables, and (iii)~produce a response for the command (i.e., the output).
%
Commands are \emph{deterministic}: the changes to the state and response of a command are a function of the state variables the command reads and the command itself.

State-machine replication provides clients with the abstraction of a highly available service while hiding the existence of multiple replicas.
This last aspect is captured by \emph{linearizability}, a consistency criterion: a system is linearizable if there is a way to reorder the client commands in a sequence that (i)~respects the semantics of the commands, as defined in their sequential specifications, and (ii)~respects the real-time ordering of commands across all clients~\cite{Attiya04}. 
In traditional state-machine replication (SMR), linearizability is achieved by having each replica execute commands sequentially in the same order.
Since commands are deterministic, replicas will produce the same state changes and response after the execution of the same command.

\begin{figure}[ht]
  \begin{center}
      \includegraphics[width=\sizefactor\columnwidth]{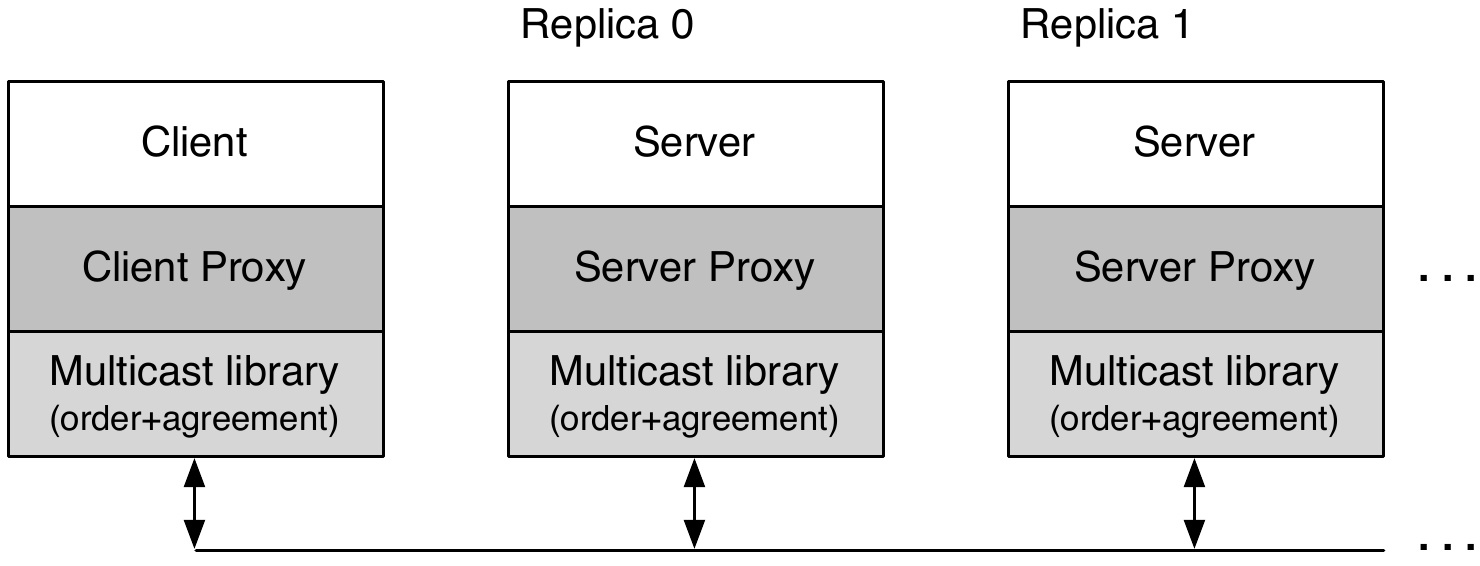} 
    \caption{A simple state-machine replication architecture.}
        \label{fig:architecture}
  \end{center}
\end{figure}

A ``commodified architecture" for state-machine replication can be organized as follows (see Figure~\ref{fig:architecture}).
%
Clients and servers interact in a way similar to remote procedure invocations~\cite{BN84}. 
Clients access the service by invoking service commands (with the appropriate parameters).
Client proxies intercept client invocations, turn them into requests that include a command identifier and the marshaled parameters, and multicast the requests to the replicas.
Requests are delivered by the server proxies, which re-assemble invocations and issue them against the local server replica.
Each server executes one command at a time.
Similarly to remote procedure calls, the client and client proxy (respectively, server and server proxy) can be implemented as a single process, sharing a common address space.
The command's response follows the reverse path to the client using one-to-one communication.
Even though the client proxy may receive the response for a command from multiple servers, all responses are the same and the proxy returns only one response to the client.

It has been observed that replicas can execute ``independent commands" concurrently without violating consistency~\cite{Sch90}.
Two commands are \emph{independent} if they either access different variables or only read variables commonly accessed; conversely, two commands are \emph{dependent} if they access one common variable $v$ and at least one of the commands changes the value of $v$.
For example, two read commands are independent, while a read and an update command on the same variable are dependent.
A few approaches have been suggested in the literature to execute independent commands concurrently with the goal of improving performance~\cite{KWQCAD2012, KD2004}.
Although these approaches differ in how they achieve concurrency, they share the fact that commands are totally ordered and each replica delivers commands sequentially as a single stream.
Since the execution of independent commands is concurrent but delivery (and possibly scheduling) is sequential, hereafter, we refer to these approaches as \emph{semi-parallel state-machine replication} (sP-SMR).


This paper proposes to parallelize both the execution and the delivery of commands, an approach we refer to as \emph{parallel state-machine replication} (P-SMR) (see Table~\ref{tbl:how}).
P-SMR uses multiple multicast groups to partially order commands across replicas, where each group leads to a different stream of commands delivered at the replica. 
%
P-SMR improves on traditional state-machine replication by allowing independent commands to execute concurrently.
It has two advantages over semi-parallel state-machine replication.
First, since a replica can handle multiple streams of commands, multicast can be implemented more efficiently.
This happens because each command stream is independent of the other, and can use different threads within a node or even involve different sets of nodes per stream.
For example, our multicast library uses one Paxos~\cite{Lam98} instance per stream, and each stream can have a different set of acceptor nodes~(see Section~\ref{sec:amcastimpl}).
Moreover, command streams can be mapped to multiple network interfaces, commonly available in modern servers.
Hence, communication at a server is not limited by the bandwidth of a single network interface, but by the aggregate bandwidth of the server's interfaces.
Second, independent commands are not delivered by a single component and then scheduled for parallel execution; they are directly delivered by multiple worker threads, thereby reducing overhead.


\begin{table}[htdp]
\centering
\begin{tabular}{l|c|c|c}
Command... \hspace{10mm}			& SMR		& sP-SMR	& P-SMR	\\ \hline
\hfill ...delivery & sequential 	& sequential	& parallel	\\
\hfill ...execution	& sequential 	& parallel		& parallel	\\
\end{tabular}
\caption{Degrees of parallelism in state-machine replication.}
\label{tbl:how}
\end{table}%

\section{Parallel State-Machine Replication}
\label{sec:psmr}

We present next P-SMR's design principles (Section~\ref{sec:designrationale}), consider architectural issues (Section~\ref {sec:clientserver}), detail the protocol (Section~\ref{sec:pdesign}), consider its advantages and limitations (Section~\ref{sec:advlim}), and 
show that P-SMR is linearizable and deadlock-free (Section~\ref{sec:correct}).

\subsection{Design rationale}
\label{sec:designrationale}

P-SMR's design is guided by two principles: \emph{(1)~optimize performance for the common case}, and \emph{(2)~keep replication transparent}. 
P-SMR targets workloads dominated by independent commands.
This is the case of services that process mostly read commands (e.g., name services) or whose state can be partitioned so that most commands access variables in one partition and rarely in multiple partitions (e.g., file systems).
In the state-machine replication architecture presented in Section~\ref{sec:background}, replication is transparent for clients: details related to communicating with multiple replicas are hidden from the clients and handled by the client proxies and the multicast library.
Similarly to SMR, P-SMR should not expose replication details to the client application.

\subsection{Client and server organization}
\label{sec:clientserver}

P-SMR builds on the architecture depicted in Figure~\ref{fig:architecture}.
Client and server proxies are created based on (a)~the \emph{signature} of each service command, including the command's identifier and a description of the command's input and output parameters with their types; and
(b)~the \emph{command dependencies (C-Dep)}, specifying which commands depend on each other.
Therefore, in addition to providing the server's code, the service designer must also provide the command signatures and the C-Dep.
Although the C-Dep can be automatically generated from the signatures and the server's code, in our prototype C-Deps were created manually.

At the clients, command signatures are used by the client proxy to create a request from the client invocations and return a response to the client.
At the servers, command signatures are used by the server proxy to turn delivered requests into local server invocations and assemble the response of commands.
In both cases, the process is analogous to the one used in the commodified state-machine replication architecture presented in Section~\ref{sec:background}.

C-Dep is used to automatically compute the \emph{Command-to-Groups (C-G)} function, used by the client proxy to determine the multicast groups a request must be multicast to and by the server proxy to coordinate the local execution of dependent commands---more details in the next section.
Similarly to SMR, a client application in P-SMR will be oblivious to replication.
Moreover, since coordination among worker threads, in the case of dependent commands, is handled by the server proxy, a server designed for state-machine replication will work unchanged in P-SMR.


\subsection{Protocol design}
\label{sec:pdesign}

P-SMR takes as input the command dependencies (C-Dep) of a service and the desired multiprogramming level (MPL) at the replicas to define how independent commands can be executed concurrently and dependent commands are synchronized.
The multiprogramming level is a parameter of the system that defines the number of worker threads at the servers. 
It can be set, for example, based on the number of processing units (i.e., cores) at the servers.
In a configuration where MPL is set to $k$, we identify worker threads as $t_1, ..., t_k$. 
P-SMR organizes threads in $k$ multicast groups such that the $i$-th thread of each replica, $t_i$, belongs to group $g_i$.

\begin{figure*}[ht]
  \begin{center}
      \includegraphics[width=\sizefactor\textwidth]{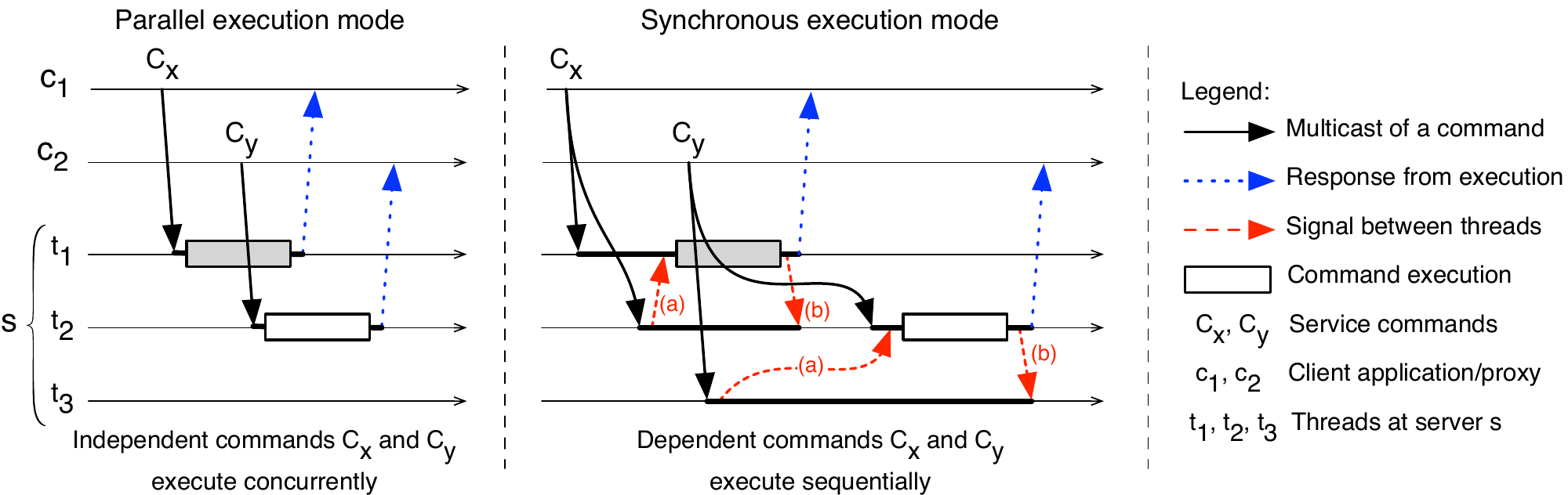} 
    \caption{Two execution modes in P-SMR, parallel (left) and synchronous (middle). For clarity, we show the execution of clients $c_1$ and $c_2$ against a single server replica $s$ with three worker threads, $t_1, t_2$ and $t_3$.}
        \label{fig:execmodel}
  \end{center}
\end{figure*}

\emph{Basic principle.}
%
A client proxy executes command $C$ by multicasting a request with $C$ to a set of destination groups, computed by the C-G function (see Algorithm~1).
Worker threads at the server proxy deliver commands and invoke their execution against the local server.
The execution of a worker thread alternates between two modes:
The thread is in \emph{parallel mode} when it delivers a command multicast to a single group and in \emph{synchronous mode} when it delivers a command multicast to multiple groups.
In parallel mode, upon delivering $C$, thread $t_i$ executes $C$, sends $C$'s response to the client and waits for the next command.
In synchronous mode, threads that deliver $C$, hereafter identified as $\tau$, synchronize using barriers:
threads in $\tau$ send a signal to one designated thread $t_i \in \tau$ (signal (a) in Figure~\ref{fig:execmodel}) and wait for a signal from $t_i$; 
after $t_i$ receives the signals it executes $C$, sends $C$'s response to the client, and signals threads in $\tau$ to continue with the next command (signal (b) in Figure~\ref{fig:execmodel}).
Note that as a consequence of our execution model, two commands will be ordered consistently across replicas if they are multicast to the same group or they are dependent.


\emph{Detailed algorithm.}
To execute command $C$, invoked by an application client (line~1 in Algorithm~1), the client proxy determines all groups $\gamma$ involved in the command using the service's C-G function (line~2) and multicasts $C$ and its input parameters to groups in $\gamma$ (line~3).
The client proxy then waits for the first response from the replicas (line~4), assigns the response received to the output parameters of $C$ (line~5), and returns to the application (line~6).
Upon delivering $C$ (line~8), thread $t_i$ at a server first determines the set of groups concerned by the command (line~9).
If $C$ was multicast to a single group, then $t_i$ continues in parallel mode: $t_i$ executes $C$ (line~12) and returns the response to the client (line~13).
If $C$ was multicast to multiple groups, then $t_i$ continues in synchronous mode and determines the thread, $t_e$, among $C$'s destinations, that will execute $C$ (line~16).
If $t_i$ is in charge of executing $C$ (lines~18--23), it waits for a signal from every other thread in $C$'s destination set (lines~18--19), executes $C$ (line~20), sends the response to the client (line~21), and then signals all other threads in $C$'s destination set to continue their execution (lines~22--23).
If $t_i$ is not in charge of $C$'s execution, it signals thread $t_e$ (line~25) and waits for $C$'s execution to complete (line~26).

\begin{algorithm}
\small
\mbox{\textbf{Algorithm 1: Parallel State-Machine Replication (P-SMR)}}
\begin{distribalgo}[1]
\vspace{-3mm}
\INDENT{\emph{A client proxy $c$ executes a call to command $C$ with \\ \hfill identifier $cid$ and $\mathit{input}$ and $\mathit{output}$ parameters as follows:}
}	\STATE $\gamma \leftarrow \textit{C-G}(cid, \mathit{input})$
	\COMMENT{$\gamma$ is the set of groups involved in C}
	\STATE multicast$(\gamma, [cid, \mathit{input}])$
	\STATE wait for first response
	\STATE $\mathit{output} \leftarrow$ response
	\STATE return
\ENDINDENT
\vspace{1.5mm}
\INDENT{\emph{Thread $t_i$ at a server proxy executes a command as follows:}}
	\INDENT{\textbf{upon} deliver$([cid, \mathit{input}])$, multicast by $c$}
		\STATE $\gamma \leftarrow \textit{C-G}(cid, \mathit{input})$
		\IF{$\gamma$ is a singleton}
			\STATE \emph{// Thread $t_i$ is in parallel mode}
			\STATE execute $cid$ with $\mathit{input}$ parameters
			\STATE send response to $c$
		\ELSE
			\STATE \emph{// Thread $t_i$ is in synchronous mode}
			\STATE $e \leftarrow min\{ j : g_j \in \gamma\}$
			\COMMENT{pick a thread deterministically}
			\IF{$i = e$}
				\INDENT{\textbf{for each} $j \neq i$ such that $g_j \in \gamma$}
					\STATE wait for signal from $t_j$
				\ENDINDENT		
				\STATE execute $cid$ with $\mathit{input}$ parameters
				\STATE send response to $c$
				\INDENT{\textbf{for each} $j \neq i$ such that $g_j \in \gamma$}
					\STATE signal $t_j$			
					\COMMENT{for thread $t_j$ to resume its execution}
				\ENDINDENT		
			\ELSE
				\STATE signal $t_e$
				\STATE wait for signal from $t_e$
				\COMMENT{thread $t_i$ pauses its execution}
			\ENDIF
		\ENDIF
	\ENDINDENT
\ENDINDENT
\end{distribalgo}
\label{psmralg}
\end{algorithm}

\emph{Defining command dependencies.}
In our prototype, C-Dep encodes two levels of dependency information:
commands that depend on each other, regardless their parameters (e.g., commands to create and delete objects) and commands that may be dependent, according to their parameters (e.g., two updates on the same object).
C-Dep includes all such interdependencies; if no entry exists in C-Dep asserting the dependency of two commands, they are independent.
Although our encoding is simple, more complex schemes could be used (e.g., \cite{KD2004}).
In Section~\ref{sec:applic} we show how this scheme can represent interdependencies in two general services, a key-value store and a networked file system.

\emph{Mapping commands to destinations.}
The client proxy determines the destination groups of a command using a Command-to-Group (C-G) function that maps the command id and its input parameters to a set of multicast groups.
The C-G is part of the client proxy; it is created based on the multiprogramming level and the service's C-Dep.
Computing the C-G requires solving an optimization problem that seeks to maximize concurrency among independent commands while keeping dependent commands synchronized. 
Allowing independent commands to execute concurrently is achieved by assigning them to different groups; 
ensuring proper synchronization amounts to assigning at least one common group to any two dependent commands.

The amount of concurrency in a service depends on the interdependencies among the service's commands.
These interdependencies are defined by the code that implements each command.
In P-SMR the interdependencies among commands are captured by the commands dependencies list (C-Dep), provided by the service designer together with the command code that runs at the replicas (see Section~\ref{sec:clientserver}).
Therefore, a C-Dep that tightly captures interdependencies will likely result in more concurrency at the replicas.
For example, consider a service with \texttt{get\_state(in: int x, out: char[] v)} and \texttt{set\_state(in: int x, char[]v)} commands, where \texttt{x} is an object identifier and \texttt{v} an object value.
A simple C-Dep would state that \texttt{set\_state} depends on any other command, regardless the object accessed.
Defining such a C-Dep requires inspecting commands \texttt{get\_state} and \texttt{set\_state} and concluding that the first reads the service's state and the second modifies the service's state.
The C-G for this C-Dep assigns a \texttt{get\_state} command to a single group and a \texttt{set\_state} command to all groups, as shown next, where \texttt{get\_state} is assigned to a random group between 1 and $k$ (recall that $k$ is the multiprogramming level):

\vspace{-1mm}
\begin{itemize}
\small
\itemsep0.08em
\item[] \hspace{5mm}\textbf{function} \textit{G-C}$(cid)$
\item[] \hspace{9mm}\textbf{switch} $(cid)$
\item[] \hspace{13mm}\textbf{case} \texttt{get\_state}: return(random$(1..k)$)
\item[] \hspace{13mm}\textbf{case} \texttt{set\_state}: return($\it{ALL\_GROUPS}$)
\end{itemize}

A more complex C-Dep identifies that \texttt{set\_state} depends only on other commands on the same object.
In this case, the C-G can assign commands on the same object to the same group and commands on different objets to different groups:

\begin{itemize}
\small
\itemsep0.08em
\item[] \hspace{5mm}\textbf{function} \textit{G-C}$(cid, x)$
\item[] \hspace{9mm}return($(x\ \textrm{mod}\ k) + 1$)
\end{itemize}

The result is that commands assigned to different groups can execute concurrently, even if they both modify the state of objects.
Moreover, additional information, if available, can be used when computing the C-G function.
For example, objects that are commonly accessed could be assigned to different groups, allowing increased concurrency.

\subsection{Advantages and limitations}
\label{sec:advlim}

In the following, we compare P-SMR to different approaches to state-machine replication according to three aspects: performance, transparency, and load balancing.
 
\emph{Performance.}
While P-SMR and sP-SMR improve the performance of state-machine replication by allowing independent commands to execute concurrently, P-SMR has two advantages with respect to sP-SMR.
First, P-SMR offloads scheduling decisions from the replicas, avoiding a bottleneck-prone scheduler, which must deliver a single stream of commands and assign them to the worker threads for execution.
This is important in time-critical services running at main-memory speed and in modern servers equipped with an ever-increasing number of processing units.
Second, since replicas in P-SMR can handle multiple parallel streams of commands, they can make better use of local hardware resources (e.g., command streams can be distributed among multiple network interfaces) and allow efficient multicast implementations, with different sets of nodes responsible for ordering different streams of commands  (see Section~\ref{sec:amcastimpl} for more details).

\emph{Transparency.}
Both P-SMR and sP-SMR require more information about a service than state-machine replication.
In both P-SMR and sP-SMR, commands that depend on each other must be identified (the C-Dep structure), although some implementations of sP-SMR~\cite{KWQCAD2012} can cope with mistakes in C-Dep (see Section~\ref{sec:rwork}).
State-machine replication does not need C-Dep since commands are executed sequentially.
In both P-SMR and sP-SMR, however, the client application is oblivious to these details, which must be identified by the service designer or automatically inferred from the server's code.
In P-SMR, the client proxy uses the C-G function, derived from C-Dep, to identify the set of groups each command must be multicast to;
in sP-SMR, the scheduler uses C-Dep to schedule independent commands concurrently and dependent commands sequentially.
Additionally, in P-SMR client and server proxies must agree on the multiprogramming level used at the servers.



\emph{Load balancing.}
In P-SMR commands are assigned to working threads \emph{statically} using the C-G function, which is computed based on the desired multiprogramming level and the command dependencies; in sP-SMR, commands are assigned to working threads \emph{dynamically} by the scheduler.
Consequently, load balancing in P-SMR is more limited than in sP-SMR.
For example, in the service described in Section~\ref{sec:pdesign}, if heavily accessed objects are assigned to the same group, then one worker thread will end up executing  more commands than the other worker threads.
If heavily accessed objects are known in advance, this information can be used when computing the C-G function so that such objects are assigned to distinct groups.
Accommodating dynamic changes in access patterns, however, would require recomputing the C-G.
While updating the C-G function online is possible, in our prototype this is done offline. 
In Section~\ref{sec:evaluation} we assess the performance of P-SMR under skewed workloads and compare the results to other approaches.

\subsection{Correctness}
\label{sec:correct}

We show that P-SMR is linearizable and deadlock-free.

\emph{P-SMR is linearizable.}
From the definition of linearizability (see Section~\ref{sec:background}), we must show that there is a permutation $\pi$ of commands in $\mathcal{E}$ that respects  (i)~the real-time ordering of commands across all clients, and (ii)~the semantics of the commands.
Let $C_x$ and $C_y$ be two commands in $\mathcal{E}$ submitted by clients $c_x$ and $c_y$, respectively. 
There are two cases to consider.

\emph{Case (a): $C_x$ and $C_y$ are independent.}
Thus, either $C_x$ and $C_y$ access disjoint sets of variables or only read variables commonly accessed.
Consequently, the execution of one command does not affect the execution of the other and they can be placed in any relative order in $\pi$.
We arrange $C_x$ and $C_y$ in $\pi$ so that their relative order respects their real-time dependencies, if any.

\emph{Case (b): $C_x$ and $C_y$ are dependent.}
Assume $C_x$ and $C_y$ are multicast to groups in $\gamma_x$ and $\gamma_y$, respectively.
From the fact that $C_x$ and $C_y$ depend on each other, $\gamma_{xy} = \gamma_x \cap \gamma_y \neq \emptyset$.
In every correct server $s$, $C_x$ (resp. $C_y$) is delivered by all threads in groups in $\gamma_x$ (resp. $\gamma_y$) and executed by one thread, say $t_x$ (resp. $t_y$).
From the order property of atomic multicast, every thread in groups in $\gamma_{xy}$ delivers $C_x$ and $C_y$ in the same relative order.
Without lack of generality, assume $C_x$ is delivered before $C_y$.

We first claim that $t_x$ executes $C_x$ before $t_y$ executes $C_y$ and the execution satisfies the sequential semantics of the commands.
To see why, notice that $t_x$ only executes $C_x$ after $t_i$ delivers $C_x$ and every other thread in groups in $\gamma_x$ delivers $C_x$ and signal $t_i$. 
Every thread $t \neq t_i$ in a group in $\gamma_x$ then waits until $t_i$ executes $C_i$ to proceed with the next command.
Thus, $t_y$ will only receive a signal from threads in groups in $\gamma_{xy}$ and execute $C_y$ after $t_x$ has executed $C_x$.
Consequently, the two commands execute in sequence, which satisfies their semantics.

We now claim that the delivery order satisfies any real-time constraints among $C_x$ and $C_y$.
Without lack of generality, assume $C_x$ finishes before $C_y$ starts, that is, $C_x$ precedes $C_y$ in real time.
Thus, before $C_y$ is multicast by a client, $C_x$ has completed (i.e,. its client has received $C_x$'s response).
The claim follows from the fact that before $C_x$ is executed, it must be multicast, and thus $C_x$ is delivered before $C_y$.

From the claims above, we can arrange $C_x$ and $C_y$ in $\pi$ according to their delivery order so that the execution of each command satisfies its semantics.


\emph{P-SMR is deadlock-free.}
For a contradiction, assume a deadlock where thread $x_1$ waits for $x_2, ..., x_l$ waits for $x_1$.
Let $p(x)$ (resp. $n(x)$) be the thread that precedes (resp. succeeds) $x$ in the deadlock chain.
Thread $x$ waits for $n(x)$ if 
(1)~there is a command $C_{x,n(x)}$ multicast to groups that contain $x$ and $n(x)$;
(2)~$x$ delivered $C_{x,n(x)}$; and
(3)~$x$ needs a signal from $n(x)$ 
(a)~before $x$ executes $C_{x,n(x)}$ or 
(b)~after $n(x)$ executes $C_{x,n(x)}$.

We now claim that $x$ delivers $C_{x,n(x)}$ before $C_{p(x),x}$, that is, $C_{x,n(x)} < C_{p(x),x}$.
To see why, assume $x$ delivers $C_{p(x),x}$ before $C_{x,n(x)}$.
From the algorithm, when $x$ delivers $C_{x,n(x)}$ it has 
(a)~sent a signal to $p(x)$, if $p(x)$ was to execute $C_{p(x),x}$ or
(b)~received a signal from $p(x)$, if $x$ was to execute $C_{p(x),x}$.
In both cases, $p(x)$ cannot wait for $x$, as assumed in our deadlock chain.

From the claim above, $C_{x_l,x_1} < C_{x_{l-1},x_l} < ... < C_{x_l,x_1}$, which contradicts the atomic multicast order property.

\section{Services}
\label{sec:applic}

In this section, we show how highly available services replicated using state-machine replication can use P-SMR. 
We consider two services, a key-value store and a networked file system.

\subsection{Key-value store}
\label{sec:kvservice}

The key-value store implements commands to read and modify an in-memory database, as presented below.
An insert includes key \texttt{k} and value \texttt{v} in the database and possibly returns an error code (e.g., out of memory).
A delete removes \texttt{k} from the database or returns an error code if the entry does not exist.
A read returns the value of \texttt{k} and an update replaces the current value of \texttt{k} with \texttt{v}.
In both cases, an error code is returned if the key does not exist.

\begin{itemize}
\itemsep0.2em
\small
\item \texttt{insert(in:int k, char[] v, out:int err)} 
\item \texttt{delete(in:int k, out:int err)}
\item \texttt{read(in:int k, out:char[] v, int err)}
\item \texttt{update(in:int k, char[] v, out:int err)}
\end{itemize}

The main key-value store's data structure is a B$^+$-tree.
While a read does not result in any changes in the tree, an update changes a single entry, the one corresponding to the provided key (if present).
Inserts and deletes may modify multiple entries, depending on the structure of the tree when the command is executed (i.e., requiring partitioning and joining tree cells).
%
Therefore, we establish the following dependencies between commands:
inserts and deletes depend on all commands;
an update on key $k$ depends on other updates on $k$, on reads on $k$, and on inserts and deletes.

\subsection{Networked File System}

The Networked File System (NetFS) implements a subset of all FUSE\footnote{http://fuse.sourceforge.net/} calls (commands), enough to manipulate files and directories, as described next.
Each command takes a set of parameters as input, including a complete file path name, and returns a sequence of bytes or possibly an error code.
For simplicity, NetFS does not support soft and hard links.

Some file system calls change the structure of the file system tree (i.e., what files and directories each directory contains). 
Besides, each file descriptor seen by a client when opening a file is mapped to a local file descriptor at each NetFS server. 
Such mapping is done with a hash table accessed by all threads, which must then synchronize. 
Therefore, the following NetFS calls depend on all calls: \texttt{create}, \texttt{mknod}, \texttt{mkdir}, \texttt{unlink}, \texttt{rmdir}, \texttt{open}, \texttt{utimens}, \texttt{release}, \texttt{opendir}, \texttt{releasedir}. 
Calls to \texttt{access}, \texttt{lstat}, \texttt{read}, \texttt{write} and \texttt{readdir} depend on all calls mentioned above and on each other if they use the same file path.

\section{Implementation}
\label{sec:imple}

In this section, we describe the implementation and configuration of P-SMR and the other approaches assessed in the evaluation.

\subsection{Atomic multicast}
\label{sec:amcastimpl}


The multicast library implements the abstraction of groups by composing multiple parallel instances of Paxos; each multicast group is mapped to one or more Paxos instances.
A message can be addressed to a single group only, not to multiple groups.
In our P-SMR prototype, each thread $t_i$ belongs to two groups: one group, $g_i$, to which no other thread in the server belongs, and one group $g_{all}$, to which every thread in each server belongs.
Threads deliver messages from multiple streams and use a deterministic merge mechanism to ensure ordered delivery~\cite{MPP2012}.
This is enough to implement both C-G functions presented in Section~\ref{sec:pdesign}.
Commands multicast to a group are batched by the group's coordinator (i.e., the coordinator in the corresponding Paxos instance) and order is established on batches of commands.
Each batch has a maximum size of 8 Kbytes.
The system was configured so that each Paxos instance uses 3 acceptors and can tolerate the failure of one acceptor.


\subsection{Key-value store}

The key-value store implements a B$^+$-tree where each entry has an 8-byte integer key, used as the tree index, and an 8-byte value.
The servers implement all commands described in Section~\ref{sec:kvservice}.
In order to generate enough load to reach maximum performance, each client maintains a window of outstanding requests that can contain up to 50 commands. 
The tree is initialized with 10 million keys on each replica and unless specified otherwise, clients select the keys uniformly.

In addition to P-SMR, we implemented a semi-parallel state-machine replication approach (sP-SMR), traditional state-machine replication (SMR), and a non-replicated architecture with a single multi-threaded server directly connected to the clients (no-rep). 
In no-rep and sP-SMR a scheduler at the server is responsible for scheduling incoming commands for execution at worker threads.

We also compare the approaches above to Berkeley DB version 5.3 (BDB), deployed as a client-server architecture.
We configured BDB to use the in-memory B-tree access method with transactions disabled and multithreading and locking enabled. 
Differently from P-SMR, sP-SMR and no-rep, 
BDB uses locks to synchronize the concurrent execution of commands.
As a result, there is no scheduler interposed between clients and server threads: each server thread receives requests through a separate socket, executes them, and responds to clients.
Except for the no-rep and BDB experiments, in which there is only one replica, the key-value store is fully replicated on two servers.

\subsection{Networked File System}


After NetFS is mounted at a client node, client calls are intercepted by FUSE and redirected to a local file system proxy, which multicasts them as requests to remote servers.
This design differs from the architecture presented in Section~\ref{sec:background} in which each client has its own proxy.
In NetFS, all clients at a node share the same client proxy. 
Since file system calls are intercepted by FUSE, client applications do not need to be linked with the client proxy to use NetFS.
At the server, incoming requests are received by the server proxy and executed against a local in-memory file system.

In P-SMR we created eight path ranges, each one assigned to a separate thread at the server, which corresponds to the number of cores available in each server node. 
The file system proxy at a client uses atomic multicast to submit requests to servers, according to the command and file path.
Nine multicast groups are used, eight of them for per-path requests, and one for serialized requests. 

In addition to P-SMR, we also implemented SMR and sP-SMR.
sP-SMR uses eight worker threads and also relies on atomic multicast to order commands. 
A single scheduler thread delivers all requests and, if they are independent, enqueues them for execution by one of the workers. 
In the case of a request requiring sequential execution, the scheduler waits for the worker threads to finish their ongoing work and then assigns the request to one worker thread.


In all cases, a request is compressed by the client and uncompressed by the worker thread that executes the request, which after executing the command compresses the response and sends it back to the client.
All implementations use lz4 compression algorithm.\footnote{http://code.google.com/p/lz4/}

\section{Evaluation}
\label{sec:evaluation}


In the following, we motivate our evaluation (Section~\ref{sec:evalratio}), describe the experimental setup (Section~\ref{sec:expsetup}), and detail our findings (Sections~\ref{exp:readonly}--\ref{sec:netfsperf}).

\subsection{Evaluation rationale}
\label{sec:evalratio}
%
We compare P-SMR to other approaches to state-machine replication (SMR and sP-SMR).
When assessing the key-value store service, we also consider two single-server architectures: a scheduler-worker server that uses a scheduling policy similar to servers in sP-SMR (no-rep) and a multithreaded server that relies on locks to synchronize the execution, without a scheduler (BDB). Our evaluation addresses the following aspects.

\begin{figure*}[ht]
  \begin{center}
    \begin{tabular}{cc}
      \includegraphics[width=\sizefactor\columnwidth]{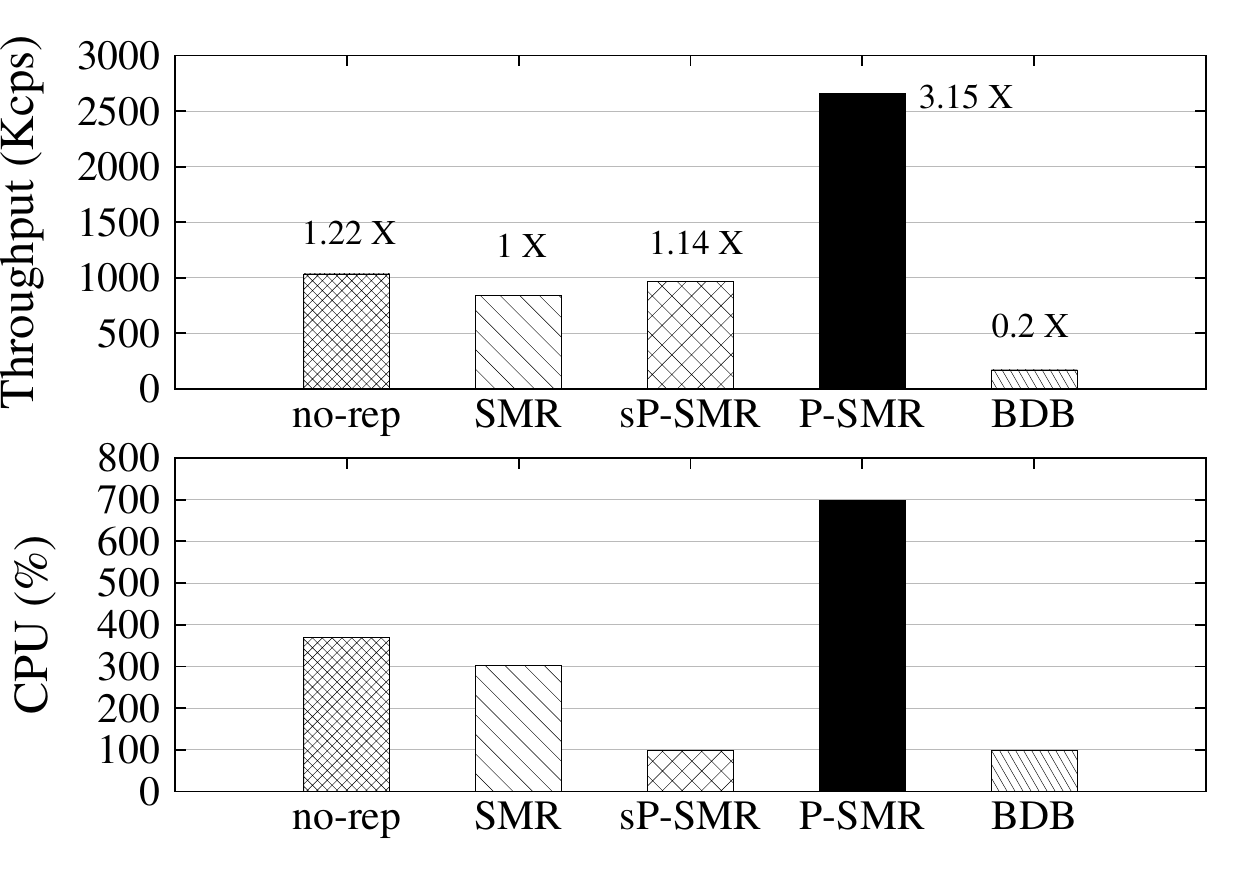} &
      \includegraphics[width=\sizefactor\columnwidth]{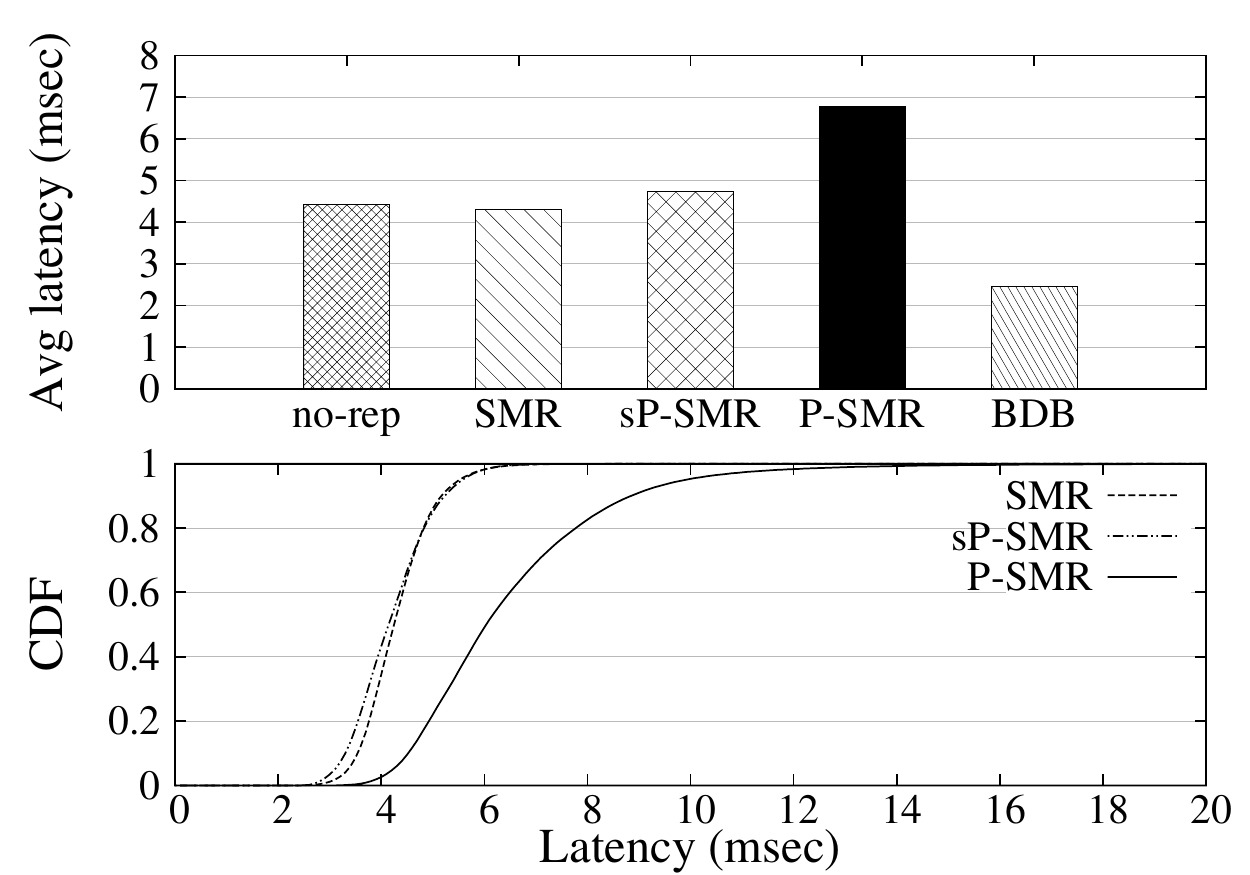}\\
    \end{tabular}
    \caption{Performance of independent commands; throughput in Kilo commands executed per second (Kcps) (top-left); CPU usage (bottom-left); average latency in milli~seconds (top-right); CDF of latency (bottom-right).}
    \label{fig:readonly}
 \end{center}
\end{figure*}

\begin{figure*}
  \begin{center}
    \begin{tabular}{cc}
      \includegraphics[width=\sizefactor\columnwidth]{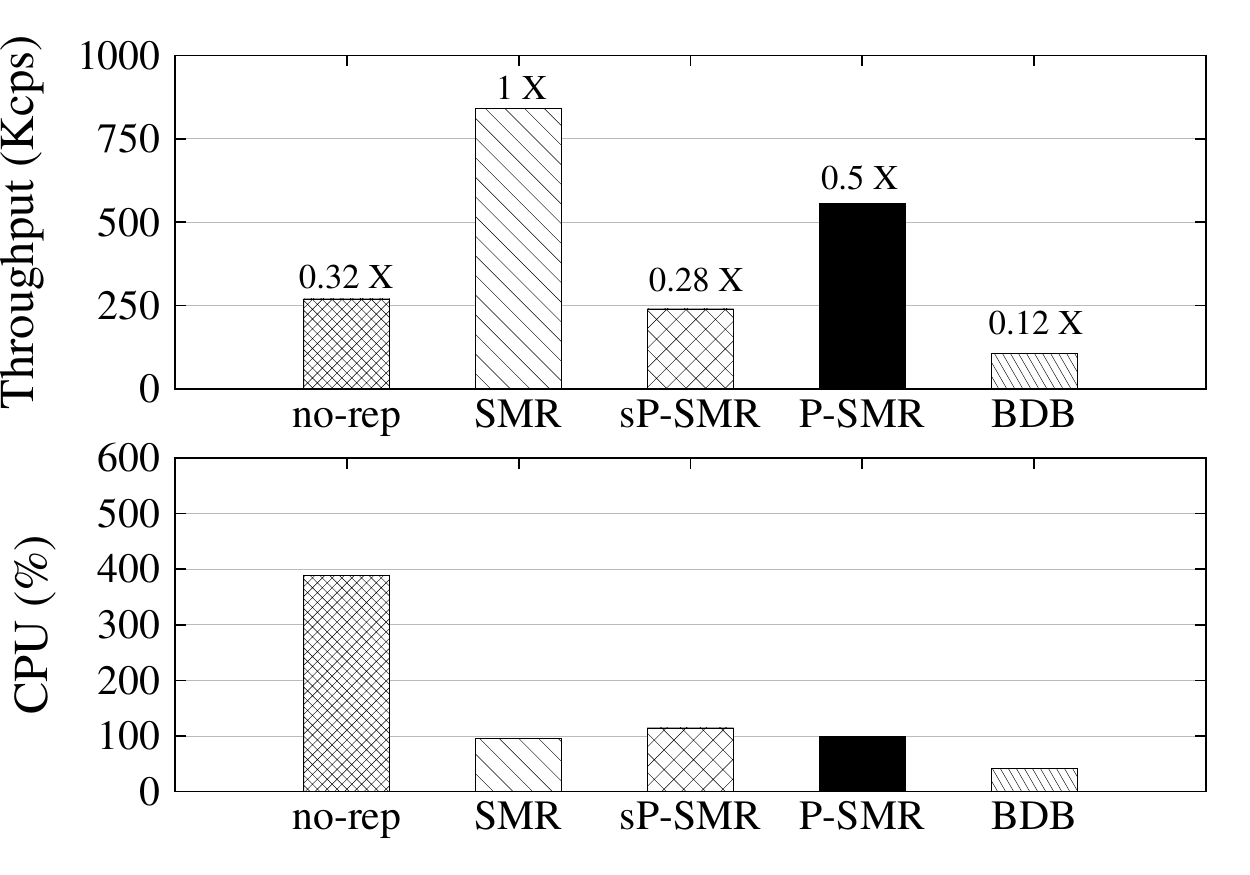}&
       \includegraphics[width=\sizefactor\columnwidth]{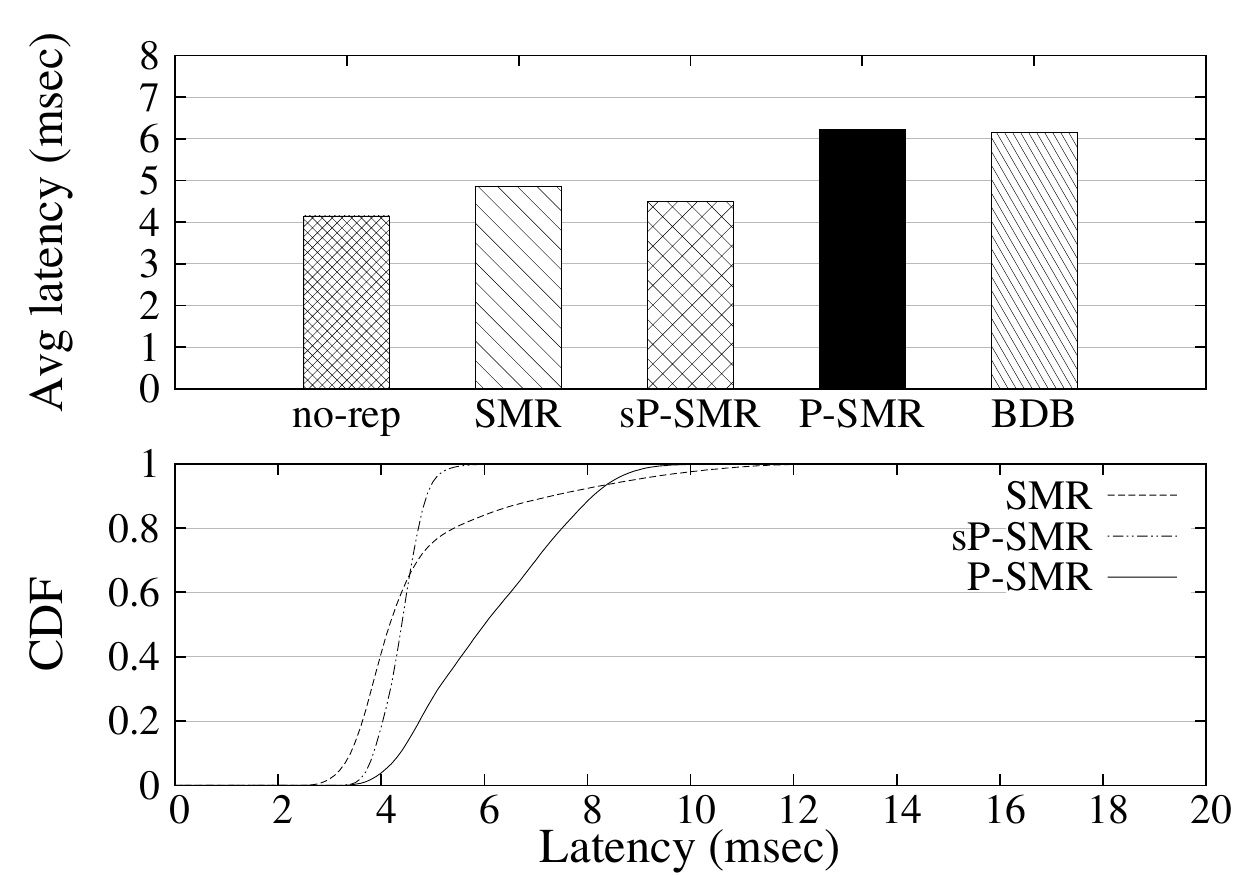}\\
    \end{tabular}
    \caption{Performance of dependent commands; throughput in Kilo commands executed per second (Kcps) (top-left); CPU usage (bottom-left); average latency in milliseconds (top-right); CDF of latency (bottom-right).}
    \label{fig:writeonly}
 \end{center}
\end{figure*}

\begin{enumerate}
\item Parallelism is only possible if commands do not contend for common variables and structures. Our first experiment (Section~\ref{exp:readonly}) compares the various techniques under workloads that allow maximum parallelism (i.e., commands are independent).

\item In the second experiment (Section~\ref{sec:perfdepcmds}), we measure the cost of synchronization in worst-case scenarios, where all commands depend on each other. 
In other words, the workload does not allow concurrency.
\item P-SMR performs best in workloads dominated by independent commands.
In this experiment (Section~\ref{sec:mix}), we seek to determine ``P-SMR's breakeven point": the percentage of dependent commands in the workload that make P-SMR neither better not worse than SMR.

\item Load balancing in P-SMR is more limited than in sP-SMR (see Section~\ref{sec:advlim}).
Consequently, in skewed workloads worker threads may end up with uneven loads, which hurts performance.
To assess the phenomenon, we compare the two techniques under skewed workloads (Section~\ref{sec:skewed}).

\item An important aspect is how the techniques scale performance with the number of threads in a server.
To answer this question, we evaluate configurations with an increasing number of worker threads and study the improvements that each thread contributes to the overall throughput (Section~\ref{exp:scal}).

\item In this experiment (Section~\ref{sec:netfsperf}), we want to understand whether our results are limited to a particular service.
We address this issue by assessing the performance of state-machine replication approaches with a Networked File System (NetFS).

\end{enumerate}

\subsection{Experimental setup}
\label{sec:expsetup}

We ran all the tests on a cluster with two types of nodes:
(a)~HP SE1102 nodes equipped with two quad-core Intel Xeon L5420 processors running at 2.5~GHz and 8~GB of main memory, and
(b)~Dell SC1435 nodes equipped with two dual-core AMD Opteron processors running at 2.0~GHz and 4~GB of main memory.
The HP nodes are connected to an HP ProCurve Switch 2910al-48G gigabit network switch, and the Dell nodes are connected to an HP ProCurve 2900-48G gigabit network switch.
Each node is equipped with two network interfaces.
The switches are interconnected via a 20~Gbps link.
The nodes ran CentOS Linux 6.2 64-bit with kernel 2.6.32.
Clients were deployed on the Dell nodes; Paxos's acceptors and servers were deployed on the HP nodes.

\subsection{Performance of independent commands}
\label{exp:readonly}


\emph{Benchmark setup:}
In this experiment we evaluate the key-value store with a workload composed of read commands only. The values we report correspond to the peak throughput of each technique and are obtained with 8 threads for P-SMR, 2 threads for sP-SMR and no-rep, 1 thread for SMR, and 6 threads for BDB. In case of norep and sP-SMR the number of threads excludes the scheduler.


\begin{figure*}
  \begin{center}
    \begin{tabular}{cc}
       \includegraphics[width=\sizefactor\columnwidth]{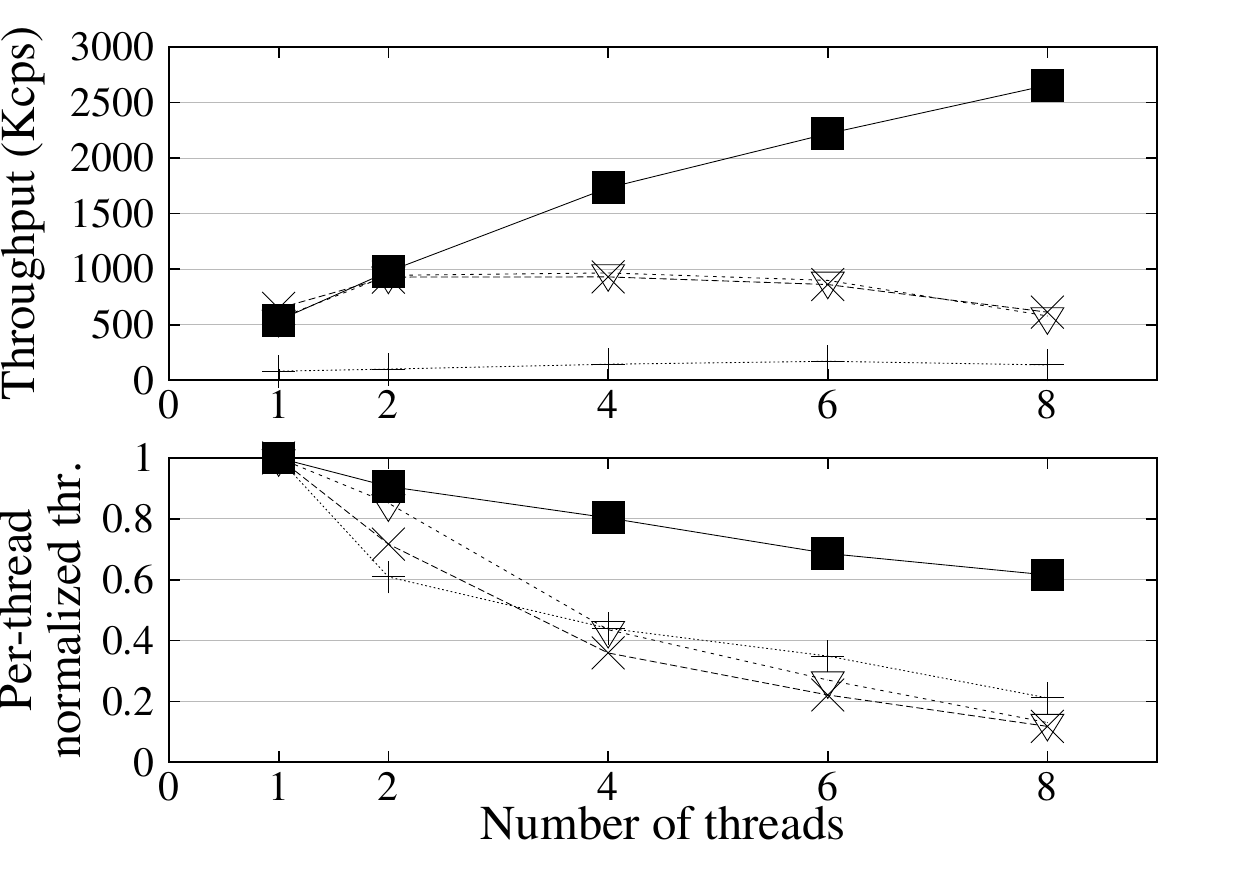}&
       \includegraphics[width=\sizefactor\columnwidth]{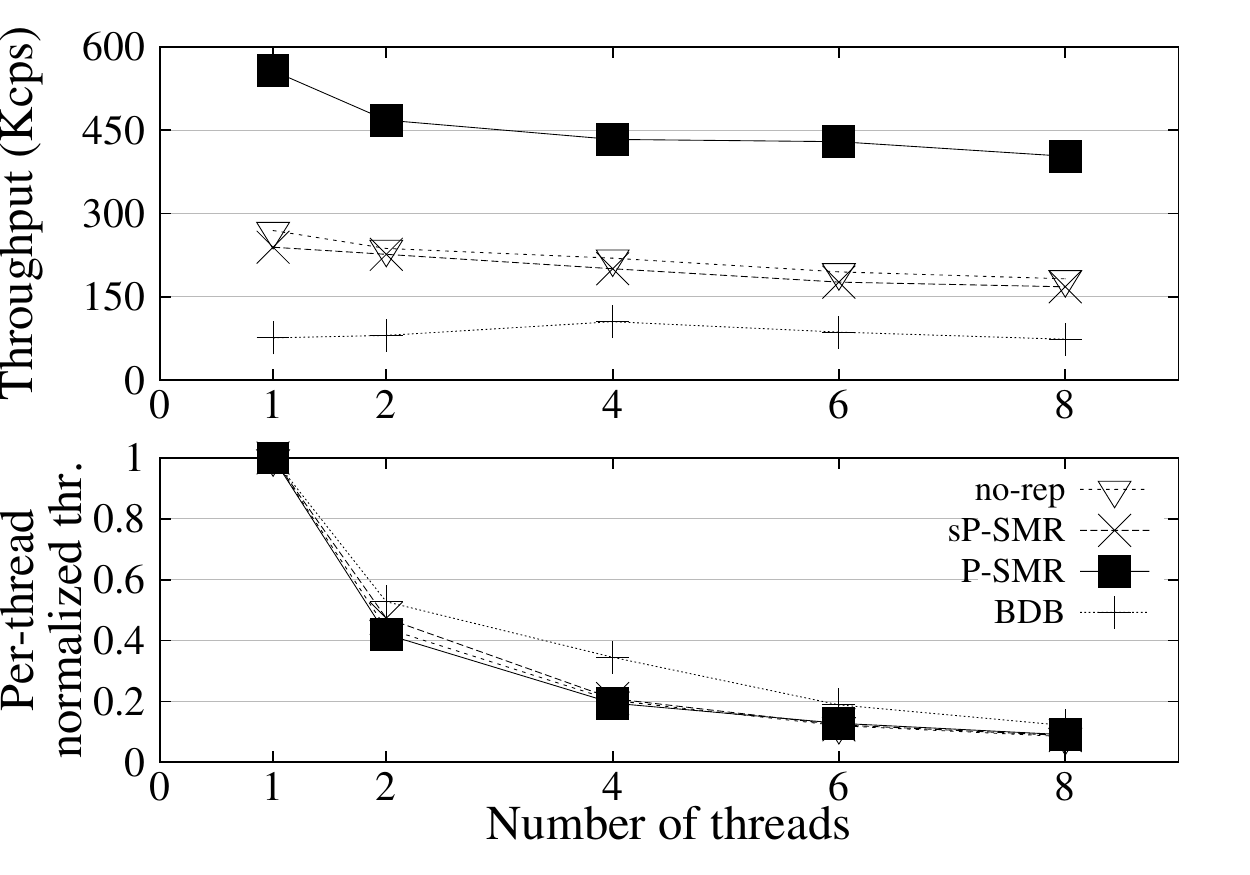}\\
    \end{tabular}
    \caption{The effect of the number of threads on the performance independent commands (left) and dependent commands (right); maximum throughput in Kilo commands executed per second (Kcps) (top graphs); normalized per-thread throughput (bottom graphs). }
    \label{fig:scalability}
 \end{center}
\end{figure*}

\emph{Results:} 
The throughput of P-SMR is about 3.15 and 2.75 times higher than SMR and sP-SMR respectively (Figure~\ref{fig:readonly}). 
The scheduler in sP-SMR and no-rep becomes CPU-bound and does not allow performance to increase. The throughput of SMR is limited by what a single thread can achieve, whereas no-rep is multithreaded and achieves higher throughput. The throughput of no-rep is slightly higher than sP-SMR as the latter is built on top of atomic multicast.
BDB has the lowest throughput due to high overhead with locking, reflected in the CPU usage. Latency of P-SMR is the highest, as it also achieves the highest peak throughput among all techniques. Although not shown, in lower throughputs P-SMR has a comparable latency to the others. no-rep's latency is slightly higher than SMR due to the overhead of the scheduler. Latency of sP-SMR is affected by both the overhead of ordering and scheduling and is higher than latency of no-rep and SMR. 

\subsection{Performance of dependent commands}
\label{sec:perfdepcmds} 


\emph{Benchmark setup:}
In this experiment we determine the maximum throughput of the key-value store service when commands are inserts and deletes. The values are obtained with 4 threads for BDB and with 1 thread for all the other techniques.
In case of norep and sP-SMR the number of threads excludes the scheduler. 

\emph{Results:} 
SMR is single-threaded and thus is not subject to synchronization overhead which consequently allows it to reach the highest throughput (Figure~\ref{fig:writeonly}).

Moreover, throughput in SMR remains constant at about 842K cps, both with independent and dependent commands; in BDB the throughput decreases from 140K cps to 105 K cps.
%
%
P-SMR's latency is higher than SMR's and sP-SMR's. The long tail in the CDF graphs suggest that P-SMR's latency is subject to more variation than SMR's and sP-SMR's.

\subsection{Scalability}
\label{exp:scal}


\emph{Benchmark setup:}  
We measure the maximum throughput of the key-value store service while the number of threads changes from one to eight when commands are independent and then dependent.
In sP-SMR, the number of threads reflects the worker threads excluding the scheduler. 
We report absolute values for the maximum throughput and the normalized throughput of an individual thread.

\emph{Results:} 
With independent commands only, the throughput of all the techniques, except for BDB, compare equally with one thread (Figure~\ref{fig:scalability}). 
As threads are added, the throughput of all the techniques, except for P-SMR, decreases. 
For sP-SMR and no-rep this happens due to scheduling overhead at the scheduler. 
P-SMR has better scalability than the other techniques (see bottom left graph). 
With dependent-only commands, in all the approaches, except BDB, throughput decreases with the number of worker threads. We attribute this to the overhead of synchronization. The throughput of BDB increases up to 4 threads and then it decreases due to locking overhead.

\begin{figure*}[ht]
\hspace{0.5cm}
\begin{minipage}[b]{0.45\linewidth} 
\centering
\includegraphics[width=0.95\columnwidth]{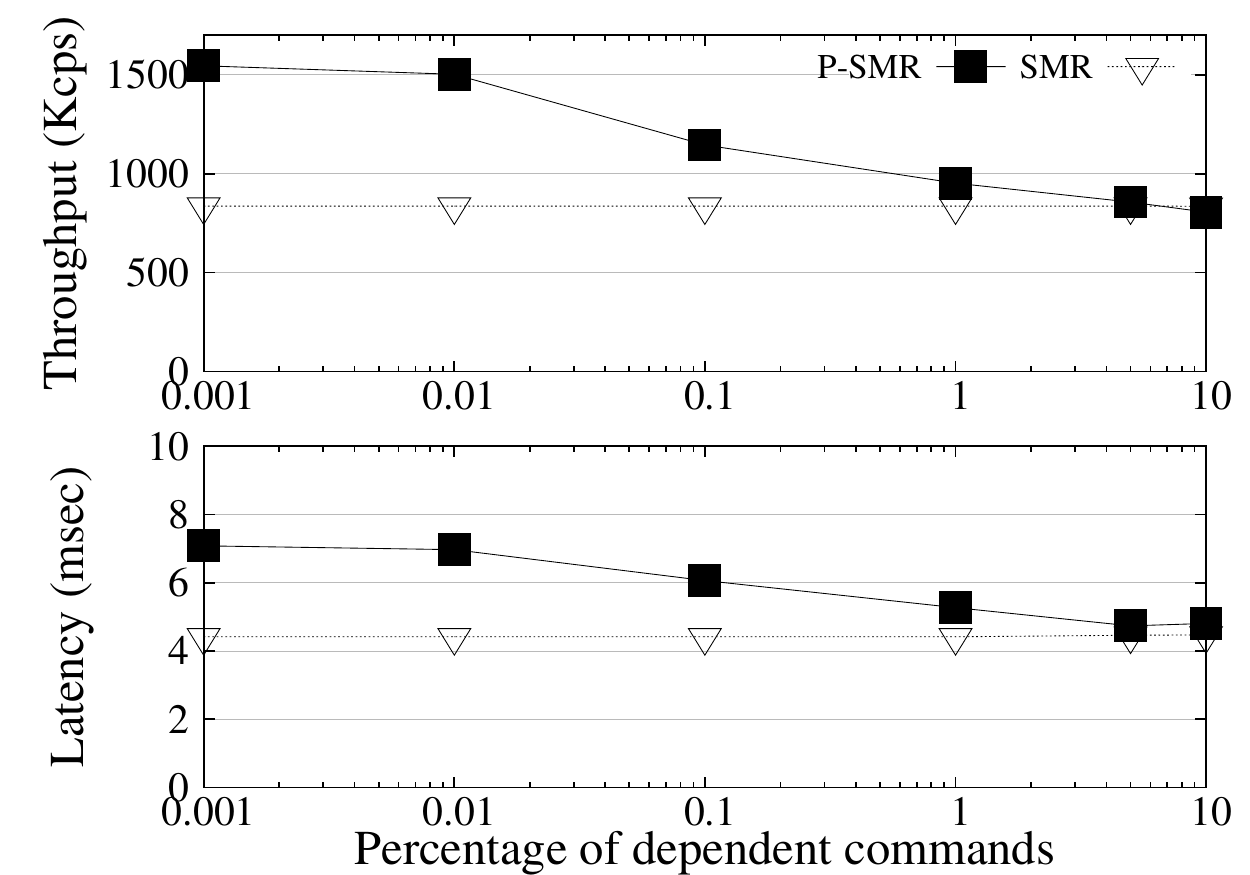}  
\caption{Performance of mixed workloads (both independent and dependent commands); throughput measured in Kilo commands executed per second (Kcps) (top); the average latency measured in milliseconds (bottom); x-axis is in log scale.}
\label{fig:mixed}
\end{minipage}
\hspace{0.1cm} 
\begin{minipage}[b]{0.45\linewidth}
\centering
\includegraphics[width=0.95\columnwidth]{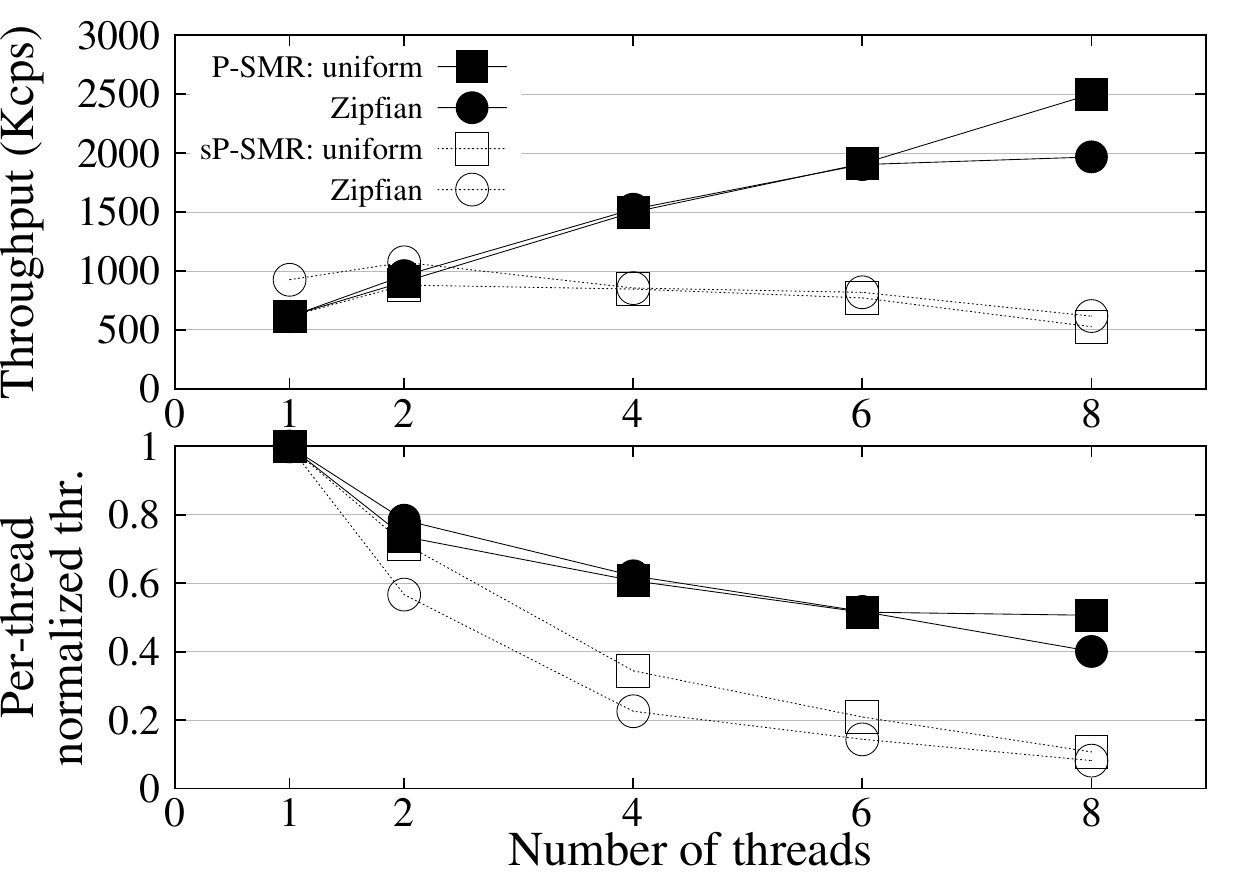}
\caption{The effect of the number of threads on the performance of a skewed workload; maximum throughput in Kilo commands executed per second (Kcps) (top); normalized per-thread throughput (bottom).}
\label{fig:zipf}
\end{minipage}
\end{figure*}


\subsection{Performance of mixed workloads}
\label{sec:mix}

\emph{Benchmark setup:} 
We measure the maximum throughput of the key-value store service with workloads composed of inserts, deletes, and reads. 
The x-axis shows the percentage of dependent commands (inserts and deletes) with respect to all the commands in the workload. 
P-SMR uses 8 workers in this experiment. 
We compare the performance of P-SMR to SMR, the only approach that is not subject to synchronization overhead and therefore has the highest performance under dependent commands.
The average latency corresponds to the maximum throughput. 

\emph{Results:} 
SMR's throughput remains constant with the workload mix (Figure~\ref{fig:mixed}).
This is expected since most of the cost to execute a read, insert and delete operation is related to traversing the tree (statistics gathering starts after the tree is initialized; thus, few inserts and deletes involve changes in multiple levels of the tree).
P-SMR's throughput is above SMR's up to about 10\% of dependent commands.
The reduction in performance is due to synchronization overhead. 
%
P-SMR's latency decreases as the percentage of dependent commands increases.
The decrease in latency corresponds to a reduction in throughput.

\subsection{Performance of skewed workloads}
\label{sec:skewed}


\emph{Benchmark setup:} 
The workload is composed of 50\% updates and 50\% reads against the key-value store. 
We evaluate the scalability of each approach with a uniform and a Zipfian key selection distribution. 
In the latter case, clients select keys following a Zipfian distribution with exponent value of one. 
In skewed distributions, communication is expected to be unevenly balanced across multicast groups.
In sP-SMR, the number of threads reflects the number of worker threads excluding the scheduler. 
Besides absolute values for the maximum throughput we also show the normalized throughput of an individual thread. 
Perfect scalability means that the throughput of each thread remains constant as worker threads are added.

\emph{Results:} 
With a uniform selection of keys, commands are evenly distributed across groups and P-SMR's throughput increases up to the capacity of each available core, up to 6 threads (Figure~\ref{fig:zipf}).
With a Zipfian distribution, however, P-SMR's throughput is bounded by the most-loaded multicast group (point with 8 threads).
sP-SMR is not bounded by a single multicast group as is P-SMR, but by the load the scheduler can handle until it becomes CPU-bound. 
Increasing the number of worker threads after two threads has a negative impact on sP-SMR's performance since the scheduler spends more time synchronizing with worker threads.
Also notice that with 1 and 2 threads the throughput of sP-SMR with a uniform workload is lower than its throughput with a Zipfian distribution. 
In the Zipfian distribution some keys are accessed more often than the others, there are higher chances that these keys are cached at the processor.
According to the normalized per thread throughput, P-SMR scales better with the number of cores than sP-SMR under both uniform and Zipfian distributions. 




\subsection{NetFS performance}
\label{sec:netfsperf}

\emph{Benchmark setup:} 
We have performed two separate experiments, one with read commands only and one with write commands only. All calls to read and write the same file are dependent, while reading and writing different files can be done in parallel. A read request has a small input parameter (i.e., the data to be read) and a large response (i.e., bytes read). A write request contains the buffer to be written as input and a small response.
Each request reads (or writes) 1024 bytes from (to) a file.

\begin{figure}[h]
\center
\includegraphics[width=\sizefactor\columnwidth]{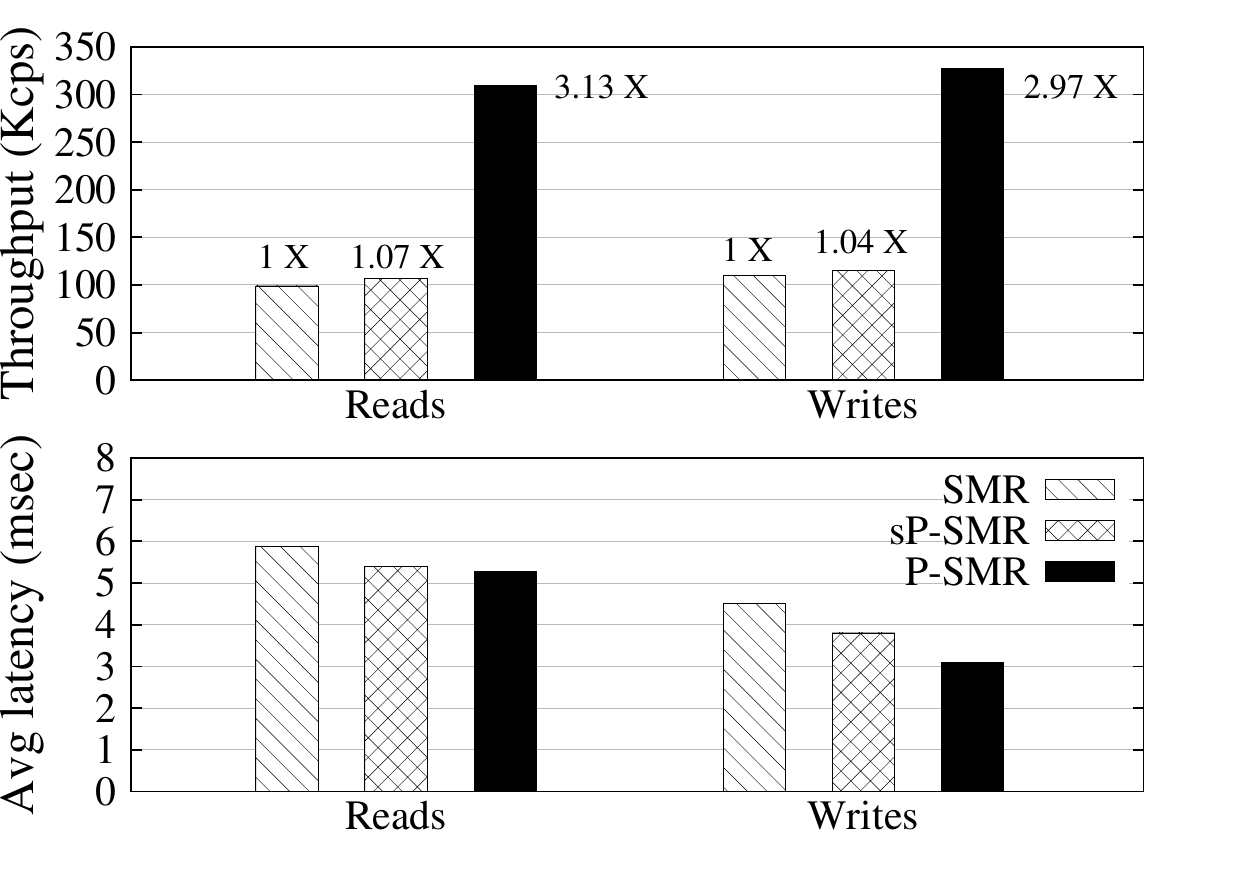}
\caption{Performance of read and write commands in NetFS; maximum throughput in Kilo commands executed per second (Kcps) (top); average latency in milli~seconds (bottom).}
\label{rnfs}
\end{figure}

\emph{Results:} 
SMR reaches a maximum throughput of approximately 100 Kcps (reads) and 110 Kcps (writes), whereas sP-SMR caps throughput at approximately 116 Kcps for both reads and writes, an improvement of about 1.2x and 1.1x, respectively (Figure~\ref{rnfs}).
The small improvement is explained by the use of CPU, where the scheduler becomes a bottleneck before fully using the remaining cores. 
P-SMR has significantly better results: read and write commands reach a maximum throughput of 309 Kcps and 327 Kcps, respectively, an improvement of 3.1x and 3x. 
%
%
Both P-SMR and sP-SMR outperform SMR, while P-SMR presents a slight advantage with respect to sP-SMR for reads and more substantial gains for writes.
Servers had to decompress requests and compress replies. As compression with lz4 takes longer than decompression, read requests took longer to execute than write requests. This is the reason for the latency difference between reads and writes.

\section{Related work}
\label{sec:rwork}


\emph{General-purpose approaches.}
Allowing multiple threads to execute commands concurrently may result in state and output inconsistencies if dependent commands are scheduled differently in two or more replicas. 
In~\cite{AWHF2010,bhcls2010, DLCOM2009, TA2010} the authors propose different approaches to enforcing deterministic multithreaded execution of commands. 
These solutions impose performance overheads and may require re-development of the service using new abstractions. 
%
%
Another solution is to allow one of the multithreaded replicas to execute commands non-deterministically and log the execution path, which will be later replayed by the rest of the replicas. 
Logging and replaying have been mainly developed for debugging and security rather than fault tolerance~\cite{AS2009, DLFC2008, PC2008, PZXYKLL2009, RB1999, VLWOCFN2011, XBH2003}.
These approaches typically have high overhead due to logging and may suffer from inaccurate replay, leading to differences among original and secondary copies. 

\emph{State-machine replication-specific approaches.}
Having replicas execute commands sequentially by a single thread does not imply that the whole replica's logic must be single-threaded.
In~\cite{SS2011}, the authors propose a staged architecture to exploit the processing power of multi-core servers. 
A replica is organized as a collection of modules connected through shared message queues. 
Although staging improves the throughput of state-machine replication, there is always only one thread sequentially executing the commands.
%
In~\cite{KD2004}, one thread at each replica delivers all commands in order and schedules them for execution against worker threads (i.e., sP-SMR model).
The scheduling of commands is deterministic and guarantees that only independent commands are executed concurrently; dependent commands are serialized and executed according to the order established upon delivery.
%
Eve~\cite{KWQCAD2012} is another instance in the sP-SMR category in which replicas agree on the internal state and output after the execution of each command rather than on their execution path. 
Eve employes an \emph{execute-verify model} by relying on speculative execution. 
On each replica, multiple threads can execute commands concurrently. 
After commands are executed, replicas verify whether they resulted in the same state changes and output.
In case of inconsistencies replicas have to rollback the commands and sequentially re-execute them in a unique agreed-upon order. 
Since replicas verify whether commands resulted in the same state changes and output, Eve can tolerate mistakes when determining command dependencies.
Although P-SMR also exploits service semantics to introduce multithreaded execution in state-machine replication, it does not rely on a single thread to deliver and schedule commands~\cite{KD2004} and to verify and possibly rollback the execution of commands~\cite{KWQCAD2012}.

\emph{Using semantics to improve performance.}
Other works have proposed the use of application semantics to improve the performance of state-machine replication (e.g., \cite{ADGF+00, PS99, MSR-TR-2005-33}).
These are based on the assumption that if two commands commute (e.g., incrementing a counter), then different replicas can execute them in different order and still reach the same final state.
These works aim at reducing the delay to deliver a command by avoiding an expensive ordering protocol when possible.

\section{Conclusions}
\label{sec:final}


State-machine replication is a fundamental approach to designing highly available services.
%
Hence, it comes as no surprise that a number of approaches have been proposed to allow multithreaded state-machine replication.
Some of these approaches take advantage of application semantics to execute independent commands concurrently and serialize dependent commands.
Parallel State-Machine Replication also uses application semantics, but differently from previous proposals, it does not rely on a single component to deliver and execute commands. 
We assessed P-SMR experimentally under several conditions and found that it outperforms classical state-machine replication by a factor of more than 3 and other approaches by a factor of more than 2.

%
%
%


\bibliographystyle{ieeetr}
\bibliography{main,multi-core-SMR-bib}



\end{document}